# Enhancement of Underwater Images with Statistical Model of Background Light and Optimization of Transmission Map

Wei Song, Yan Wang, Dongmei Huang, Antonio Liotta, Cristian Perra

*Abstract*—Underwater images often have severe quality degradation and distortion due to light absorption and scattering in the water medium. A hazed image formation model is widely used to restore the image quality. It depends on two optical parameters: the background light (BL) and the transmission map (TM). Underwater images can also be enhanced by color and contrast correction from the perspective of image processing. In this paper, we propose an effective underwater image enhancement method for underwater images in composition of underwater image restoration and color correction. Firstly, a manually annotated background lights (MABLs) database is developed. With reference to the relationship between MABLs and the histogram distributions of various underwater images, robust statistical models of BLs estimation are provided. Next, the TM of R channel is roughly estimated based on the new underwater dark channel prior (NUDCP) via the statistic of clear and high resolution (HD) underwater images, then a scene depth map based on the underwater light attenuation prior (ULAP) and an adjusted reversed saturation map (ARSM) are applied to compensate and modify the coarse TM of R channel. Next, TMs of G-B channels are estimated based on the difference of attenuation ratios between R channel and G-B channels. Finally, to improve the color and contrast of the restored image with a dehazed and natural appearance, a variation of white balance is introduced as post-processing. In order to guide the priority of underwater image enhancement, sufficient evaluations are conducted to discuss the impacts of the key parameters including BL and TM, and the importance of the color correction. Comparisons with other state-of-the-art methods demonstrate that our proposed underwater image enhancement method can achieve higher accuracy of estimated BLs, less computation time, more superior performance, and more valuable information retention.

## I. INTRODUCTION

Underwater image quality enhancement is a research area fundamental for improving the quality of experience (QoE) in advanced marine applications and services. Scene understanding, computer vision, image/video compression and transmission, underwater surveillance, are some applications and services strongly depending on the availability of high-quality input images for addressing professional and consumer expectations concerning QoE.

During media acquisition in the air, poor and varying illumination conditions can drastically change image contrast and visibility [1]. The quality degradation is even higher for underwater media acquisition considering the additional physical complexity of the water compared to the air. In particular, images restoration and enhancement methods are challenging due to the complex underwater environment where images are degraded by the influence of water turbidity, light absorption, and scattering [2].

Several methods have been proposed to measure and enhance the perceived QoE, and in particular blind image quality assessment methods are of much interest since they do not require prior knowledge of the original content and can be easily deployed in practical visual communication systems [3].

Under the water, the longer the wavelength is, the faster the light disappears. Compared with G-B lights, the red light is the most affected, so that underwater images often appear greenbluish tone. Jaffe-McGlamery et al. [4], [5] proposed a model of underwater imaging, which can be represented as a linear superposition of a direct component, a forward scattering component, and a back scattering component [6]. Hence the interactions among the light, the transmission medium and the scene can produce the fuzzy images.

Underwater images are enhanced and/or restored mainly by two kinds of algorithms and/or techniques which include image-based methods and physics-based methods. The former methods modify image pixel values via image processing to improve the contrast and brightness of hazed images. Traditional image enhancement methods (e.g., White Balance, Histogram Equalization (HE), Contrast Limited Adaptive Histogram Equalization (CLAHE) [7] and improved method [8]) can improve the visibility, color and natural appearance of outdoor terrestrial images. Yet, these methods are rarely effective for underwater images with complicated physical properties. Iqbal et al. proposed the integrated color model (ICM) [9] and the unsupervised color correction method (UCM) [10] based on histogram stretching in RGB and HSI color model to enhance the contrast and color of the image. Ancuti et al. [11] proposed the fusion-based images enhancement method, which is focused on color and contrast treatment. Ghani et al. [12]–[14] improved the ICM method by stretching the input image based on the Rayleigh distribution to preserve the details of the enhanced areas.

Physics-based methods restore underwater images by considering the basic physics of light propagation in the water medium and the theory of underwater imaging. The purpose of restoration is to deduce the parameters of the physical model and then recover the underwater images by reserving compensation processing. Due to the hazing effect of underwater images caused by light degradation and scattering, a phenomenon is similar to the effect of heavy fog in the air, He's dark channel prior (DCP) dehazing method [15] or its variations are widely used in underwater image restoration [16], [17]. Chao et al. [16] directly used the DCP to recover the underwater images, but results show a limited improvement. Chiang et al. [17] proposed wavelength compensation and image dehazing (WCID) to remove the artificial light (AL), compensate the attenuation of each channel by the wavelength, and eliminate the effect of the haze by the DCP. Considering the poor restoration of the DCP based on RGB channels ($DCP_{rgb}$), underwater DCP based on GB channels ($DCP_{gb}$) was proposed by Drews et al. [18], [19] to eliminate the effect of red channel in the underwater image. Using the $DCP_{gb}$ to obtain the transmission map could reduce the error of depth estimation. Galdran et al. [20] proposed a variant of the DCP which used the minimum operation of the inverted R channel and G-B channels to recover the images. Li et al. [21] estimated the background light by mapping the maximum intensity prior to dehaze blue-green channels and used Gray-World assumption theory to correct the red channel. Li et al. [ Underwater image restoration based on minimum information loss principle and optical properties of underwater imaging] proposed the minimal information loss principal (MILP) to attempt to estimate an optimal medium transmission map to restore underwater images. Peng et al. [22] considered object blurriness to estimate transmission map and scene depth. They further presented an improved method based on the image blurriness and light absorption to estimate more accurate background light and underwater scene depth to restore a precise color image [23]. Carlevaris et al. [24] simplified the estimation of the transmission map according to the difference between maximum values of the R channel and GB channels. The above DCP-based, variations of DCP-based, MILP-based and MIP-based restoration methods using the image formation model (IFM) are not competent to the estimation of BLs and TMs under complicated underwater lighting conditions, surrounding environments and color tones. In short, physics-based methods aim to acquire more accurate TM and BL to recover underwater images.

With the development of deep learning in the image restoration and enhancement, we have seen a shift from models that are completely designed by humans by optimization of parameters selection to systems that are trained by computers using example date from which feature vectors are extracted. Thus, learning-based methods for underwater image enhancement have been taken into consideration in recent years. Liu et al. [25] proposed the deep sparse non-negative matrix factorization (DSNMF) to estimate the image illumination to achieve image color constancy. Ding et al. [26] estimated the depth map using the Convolutional Neural Network (CNN) based on the balanced images that were produced by adaptive color correction. In order to restore underwater images, Cao et al. [27]employed a 5-layer convolutional neural network to estimate background light and used a multi-scale architecture, stacking two deep networks, a coarse global network and a refined network, in reference with [28] to predict scene depth map. Although the above methods based on the deep learning can estimate some correct BLs or depth maps and even restore underwater images, these trained network models can only adapt to some limited kinds of underwater images due to the mixed quality of synthetic underwater images. Deep learning methods are also extraordinarily time consuming compared with the physical or non-physical models under the same processing circumstance.

It is challenging to effectively restore different kinds of underwater images under diverse scenarios or/and with different distortions. Existing methods of underwater image restoration either produce inaccurate estimation of the parameters of the background light (BL) intensity and transmission maps (depth maps), or have high complexity. For example, white objects in the image, floating particles in the foreground, AL in the background region, or dim background light, can easily interrupt the correct estimation of BLs and TMs based on present state-of-the-art restoration methods. Ignoring light selective attenuation also impacts the estimation accuracy of BL and TM. In this paper, to guarantee the robustness of the proposed restoration based on IFM, we propose an effective method, including an accurate yet time-saving BL estimation model and TM estimation based on NUDCP and the optimizer of the depth map and ARSM. The color correction (CC) is regarded as a key post-processing to enhance the contrast and visibility of the restored images. The contributions of this paper are five-fold:

a) A MABLs database is established with 500 underwater images. The manual annotation was based on the general concept of "background light", that is, the light used to illuminate the background area. In order to guarantee the availability, MABLs are selected from one thousand BLs estimated manually after removing some unreliable BLs with latent error or uncertain elements. Finally, the database has proved to be highly accurate in terms of the good recovered quality of various underwater images using the MABLs. To the best of our knowledge, this is the first database for underwater image BL estimation.

b) A novel model of the background light estimation is proposed based on the statistical analysis of the distribution characteristics of each R-G-B channel, which is built on the combination of the average value, the median value and the standard deviation of each channel distribution. Compared with other recent BL estimation models, our statistical model can improve the accuracy and be significantly time-saving without any prior information about the underwater images.

c) An optimal model of TM estimation includes TM of R channel derived based on a NUDCP conforming to the distribution characteristics of HD underwater images combining with compensation of the depth map and optimization of the ARSM, and TMs of GB channels considering the obvious difference of attenuation rates between R channel and G-B channels in underwater environment. The proposed TM model can achieve superior results to the state-of-the-art models but with less complexity.

d) An enhancement method for different underwater images is given in two steps: (i) image restoration using the IFM model with the proposed BL estimator and TM optimizer; (ii) image color correction with modified white balance algorithm. This method successfully improves the quality of underwater images by adequately taking advantages of both the physics-based method and the image processing method.

e) Comprehensive experiments and assessments are conducted in this paper to deliver fair and sufficient evaluations and discussions of its impacts of the key parameters BL and TM, and underwater images enhancement quality involving other state-of-the-art methods.

The paper is organized as follows. Section II presents the related work. The proposed method is described in Section III. The obtained results are reported in Section IV. Section V presents the discussion. The conclusions of the paper are drawn in Section VI.

## II. RELATED WORK

### A. Image Formation Model

A simplified image formation model (IFM) [20], [29]–[32] is often used to approximate the propagation equation of underwater scattering in the background light, can be shown as:

$$I^c(x) = J^c(x)t^c(x) + (1-t^c(x))B^c, c \in \{r,g,b\} \quad (1)$$

where the $c$ represents one of red, green and blue color channels, $I^c(x)$ and $J^c(x)$ are the hazed intensity and restored radiance in one channel $c$ of the pixel point $x$, respectively; $B^c$ represents the intensity of the global background light (BL), i.e., three values corresponding to R-G-B channels; The values of $t^c(x) \in (0,1)$ for an entire image can form a transmission map (TM), which describes the portion of the scene radiance that is not scattered or absorbed and reaches the camera. The $t^c(x)$ can be also expressed as an exponential decay function relation to the scene depth $d(x)$ and the spectral volume attenuation coefficient $\beta(x)$ [33] :

$$t^c(x) = e^{-\beta(x)d(x)} = Nrer(c)^{d(x)}, c \in \{r,g,b\} \quad (2)$$

where $e^{-\beta(x)}$ can be represented as the normalized residual energy ratio $Nrer(c)$, which depends on the wavelength of one channel and the water type. But approximately 98% of the world's clear oceanic or coastal water (ocean type I) follows the rule [RADIATIVE TRANSFER IN THE OCEAN], where the accredited ranges of $Nrer(c)$ in R-G-B lights are 80%~85%, 93%~97%, and 95%~99%, respectively.

### B. BL Estimation

In order to facilitate the description, throughout this paper, $c$ and $c'$ represent one of RGB channels and GB channels, and $\Omega$ denotes a local patch size of $9 \times 9$ pixels.

The simplest method of the background light (BL) estimation is based on the brightest pixel in the whole underwater image. It is often not applicable to the scenarios where the foreground objects are brighter than the global background light. To reduce the impact of suspended particles in the image, $DCP_{rgb}$ based on methods [16], [17] choose the pixel located at the brightest point in the dark channel of the image to estimate BLs:

$$B^c = I^c\left(arg\ max\left(\min_{y\in\Omega}\left(\min_c I^c(y)\right)\right)\right) \quad (3)$$

To eliminate the effect of the red channel, the BLs are selected by finding the brightest pixel in $DCP_{gb}$ [18], where $I^c(y)$ in the Eq. (3) is replaced with $I^{c'}(y)$. Meanwhile the $B^c$ is selected as the brightest pixel or the average value in the input among top 0.1% brightest pixels in $DCP_{rgb}$ [34] or $DCP_{gb}$ [20].

The BLs can also be estimated by selecting the maximum difference of R and G-B channels in the input image, considering the fact that red channel attenuates much faster than green and blue channels in underwater [29]:

$$B^c = I^c\left(arg\ max\left|\max_{y\in\Omega} I^r(y) - \max_{y\in\Omega} I^{c'}(y)\right|\right) \quad (4)$$

A quad-tree subdivision algorithm was proposed to estimate the BL based on the higher score of the average pixel value subtracted by the standard deviation within the pure image region [21], [35]. This algorithm firstly divides the input image into four rectangular region, and then searches for the flat background region with highest score, and Eq. (4) is used to estimate a final BL.

A more complex BLs estimation is based on multiple BL candidates selection [16]. Three BL candidates from the top 0.1% blurry pixels in the input image, the lowest variance region and the largest blurriness region [23]. To compute the blurriness, an initial image blurriness map $P_{init}$ is computed as:

$$P_{init}(x) = \frac{1}{n}\sum_{1}^{n}|I_g(x) - Gau^{r_i,r_i}(x)| \quad (5)$$

where $I_g$ is the grayscale version of the input image $I^c$, $r_i = 2^i n + 1$ in the $k \times k$ spatial Gaussian filter $Gau^{r_i,r_i}(x)$ with variance $\sigma^2$ and $n$ is set to 4. Next, the maximum filter is used to calculate the rough blurriness map $P_r$ as:

$$P_r = \max_{y\in\Omega} P_{init}(y) \quad (6)$$

where $\Omega$ is set as the size of $5 \times 5$ suitable to the test image in the size of $400 \times 600$ pixels, The $P_r$ is refined by filling the holes caused by flat regions in the objects using morphological reconstruction, and the guided filter [Guided Image Filtering, Reference] is applied for smoothing to generate a refined blurriness maps $P_{blr}$ as:

$$P_{blr}(x) = F_g\{C_r[P_r(x)]\} \quad (7)$$

From the blurriness map, the BL candidates are determined, of which the regions are obtained by iteratively dividing the input image into four same-size regions using quad-tree decomposition. Because the image under sufficient light has a brighter BL, the weighted combination of the maximum and minimum candidate BLs is used to acquire the final BL, as

$$B^c = \alpha \times B^c_{max} + (1-\alpha) \times B^c_{min} \quad (8)$$

where $\alpha$ is selective coefficient, $B^c_{max}$ and $B^c_{min}$ are the maximum and minimum candidate BLs, respectively.

*C. TM Estimation*

The DCP was firstly proposed by He et al. [15], who considered that in most of the non-sky patches, at least one pixel of RGB channels in the local patch has an extraordinarily-low intensity (almost zero) on the statistical prior of outdoor haze-free images, described as:

$$J_{dark}^{rgb}(x) = \min_{y \in \Omega} \{\min_c J^c(y)\} = 0 \quad (9)$$

Applying the minimum filter to both sides of the Eq. (1) divided by $B^c$ as follows:

$$\min_{y \in \Omega} \left\{\min_c \frac{I^c(y)}{B^c}\right\} = \min_{y \in \Omega} \{\min_c J^c(y)\} + 1 - t_{DCP}(x) \quad (10)$$

Substituting Eq. (9) into Eq. (10), $t_{DCP}(x)$ is shown as:

$$t_{DCP}(x) = 1 - \min_{y \in \Omega} \left\{\min_c \frac{I^c(y)}{B^c}\right\} \quad (11)$$

Due to the aggressive attenuation of the red channel in the underwater image, the $DCP_{gb}$ only considers G-B channels. The $t_{UDCP}(x)$ [18] can be obtained by replacing $I^c(y)$ in the Eq. (11) with $I^{c'}(y)$. The maximum intensity prior (MIP) [24], calculating the difference between the maximum intensity of the R channel and that of GB channels, is used to easily estimate the transmission map is described as the following:

$$\begin{cases} D_{mip}(x) = \max_{y \in \Omega} I^r(y) - \max_{y \in \Omega}(I^{c'}(y)) \\ t_{MIP}(x) = D_{mip}(x) + 1 - max(D_{mip}(x)) \end{cases} \quad (12)$$

Different from the above methods, the scene depth is estimated by combining stretched three depth maps including the maximum filter of red channel $d_R$, the maximum intensity $d_D$ and the image blurriness $d_B$.

$$d_n = \theta_b[\theta_a d_D + (1 - \theta_a)d_R] + (1 - \theta_b)d_B \quad (13)$$

where $\theta_a = s(avg_c(B^c), 0.5)$ and $\theta_b = s(avg_c(I^r), 0.1)$ are determined by a similar sigmoid function. After the relative depth map is refined by the guided filter [Guided Image Filtering, Reference], the final scene depth $d_f$ is obtained by transforming the relative distance to the actual distance. The TM for the red channel is calculated as:

$$t^r(x) = e^{-\beta^r d_f} \quad (14)$$

where $\beta^r \in \left(\frac{1}{8}, \frac{1}{5}\right)$ following to [17], [36], and the TMs of G-B channels are obtained considering the attenuation ratios of G-B channels with respect to R channel [33].

## III. OUR PROPOSED METHOD

Our method works in RGB color space, aiming to not only recover the underlying scene radiance but also improve the contrast, color and visibility. The flowchart of our proposed method is shown in Fig. 1. We firstly proposed an effective statistical model to estimate the overall background lights (BLs) of RGB channels. Then, we estimate the transmission maps (TMs) of RGB channels based on the NUDCP integrated with TM optimizer, and the exponential decay function of three channels. Applying the estimated BLs and TMs into the underwater IFM, we gain a dehazed image. Finally, we further correct the brightness and color of the restored image via a simple color correction with white balancing using optimal gain factor.

*A. BL Estimation Model*

The existing BL estimation methods based on DCP or its variations [15], [18], [34], the hierarchical searching method [35], [14] or the BL candidate selection [23], have relatively high computation complexity and perform well for some specific underwater scenes. In this paper, according to BL selection mechanism, we create a first and effective manually annotated background lights database (called the labels in the training model). To seek suitable models for BL estimation of the respective R-G-B channels, we proposed linear and non-linear models for BL estimation of R channel and G-B channels. And the effectiveness of proposed MABLs and the accuracy of our proposed BL estimation model are expressed in Section V.

1) Manually annotated background lights (MABLs): Despite of many BL estimation methods, no one has provided objective assessment to the accuracy of BL estimation due to the lack of the "accurate" BLs. In this paper, we establish a first dataset of BLs for underwater images, which will serve as an important benchmark to support the related research. The dataset consists of manually annotated background lights values for 500 underwater images, hereinafter referred to as MABLs. Firstly, we collect over 2000 underwater images from the papers of [9]-[14], [25]-[17], [3], [6], [16], [10], [23]–[25] and from the YouTube, Google Images and Flickr. These underwater images are manually resized to the uniform $400 \times 600$ piexls.

Among these images, to ensure the diversity of the database, we randomly select 1000 images based on the following criteria: (1) various underwater scenes, including single fish, shoal of fish, coral, diving, and underwater archaeology, and (2) different distortions such as deep-water, low-visibility, thickly hazed, greenish-bluish, noise, turbid scenes. Around 50% of samples are discarded during the process of validity measurement, these guaranteed MABLs are retained as the final MABLs database. Although our database with 500 underwater images is limited, it is still significantly valuable, given there is no public database of underwater images unlike databases of outdoor images.

When annotating the BLs, we invited 15 participants (10 males and 5 females) from Shanghai Ocean University, who are in their twenties, non-experts and have a proper understanding on the visual images.

For 2000 hazed and coarse underwater images, the subject was asked to select the position of background light and record the corresponding pixel values on each underwater image in compliance with the principle of choosing the far scene point with high intensity. The principle represents the general concept

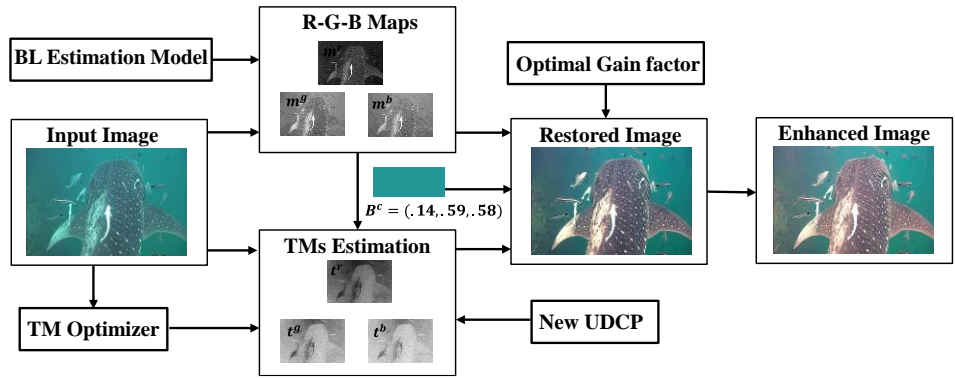

Fig. 1: The flowchart of our proposed method

of "background light": the light used to illuminate the background area. Then, from the twenty selected points of each image, we chose the one with the largest difference between the R and GB intensities and get its intensity values as the final annotated BLs. This follows the principle of underwater optical imaging: because the R light attenuates much faster than the G and B lights in the water, the difference between R and GB should be bigger when the light travels from the far point to the camera. Finally, we employed five professionals with the strong background of the image/video processing or computer vision to assess the above estimated BLs by our participants and give some valuable suggestions about improvement of the MABLs

We found the feasible rules to manually estimate the BLs, but participants and even the professionals failed to annotate BLs of existing one third of underwater images under complicated environments, such as a close scene obtained by the camera. Hence we firstly discorded those images, which cannot be annotated manually. Abiding by the rule of the minority subordinated to the majority, then we filter out candidates for which less three professionals approved of the selected BLs by the participants. Finally, we selected 500 underwater images with reliable BLs and from different types as the MABLs, and split the dataset into training and testing data in the ratio of 7:3. Although no ground truth of BLs to compare, there are some samples of the underwater images and the corresponding MABLs, which are normalized to the range of [0, 1] in Fig. 2.

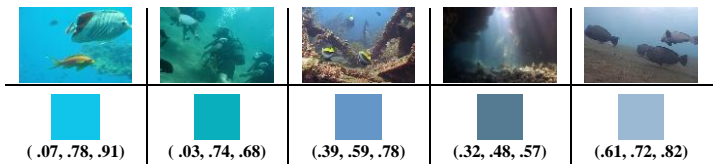

( .07, .78, .91)  ( .03, .74, .68)  (.39, .59, .78)  (.32, .48, .57)  (.61, .72, .82)

Fig. 2: Some samples of the MABLs

2) *Statistical model of BL estimation:* Based on the training data of MABLs, we propose a simple but effective model of BL estimation via statistical analysis. We have discovered the tight correlations between the MABLs values and the histogram distribution characteristics of the underwater images in different RGB color channels. Fig. 3 gives examples of five typical underwater scenes to show how the trends of histogram vary within different BLs. The red, green and blue lines in the histograms correspond to the probability distributions of R, G, and B channels, respectively.

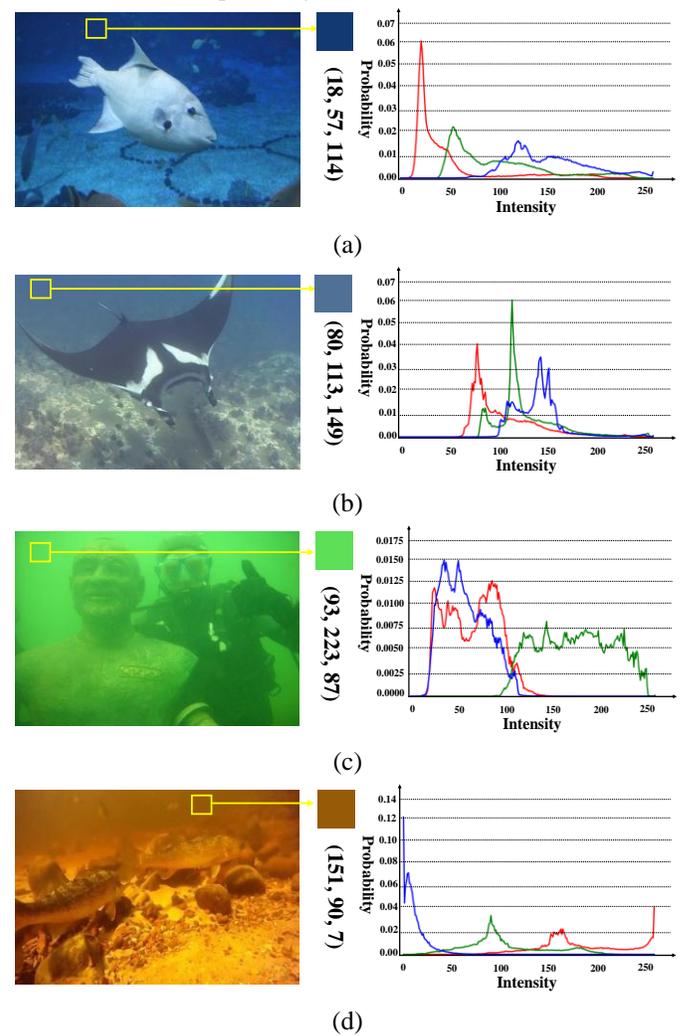

(a)

(b)

(c)

(d)

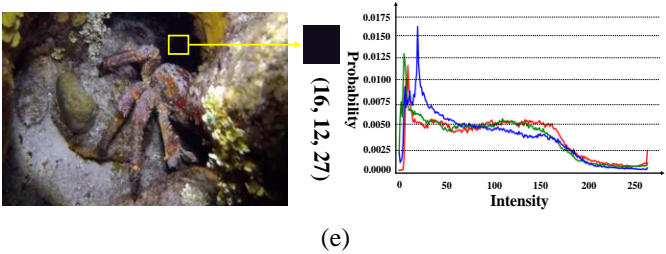

(e)

Fig. 3: Underwater images with five typical underwater scenes and the tight relationship between MABLs and the histogram distribution.

For instance, in Fig. 3 (a) and (b), the median values of R-G-B channels in the histogram are very close to the MABLs of the images. They are not affected by the large white regions in the front of the image, which can be misestimated as the BL candidate by the DCP-based, UDCP-based BL estimation algorithms that consider the brightest region under the dark channel maps of RGB channels or GB channels as the background area. In Fig. 3 (c), the green is distributed with much higher intensities than the red and blue. While in Fig. 3 (d), the red dominates the background light. Fig. 3 (e) has a near-black background light and its R-G-B distribution is consistent. To seek for the relationship between the MABLs and the characteristics of R-G-B histogram distribution, we investigated the following parameters: the average value (Avg), the median value (Med), the maximum value (Max), the minimum value (Min) and the standard deviation (Std), which fully express the relatively-concentrated distribution components in each color channel of an underwater image. We can find the distinct relationship between the MABLs and some parameters in the Fig. 4.

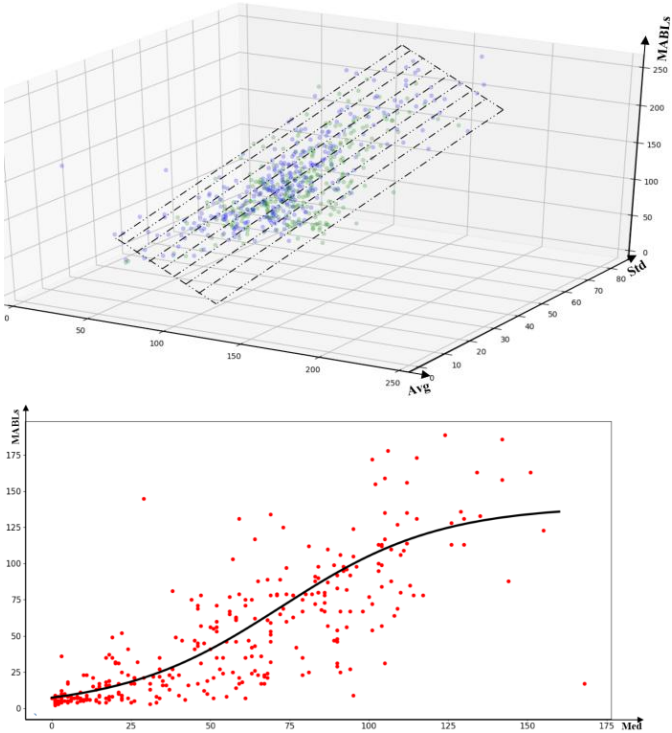

Fig. 4: The linear and non-linear regression analysis.

Pearson correlation analysis between these parameters and MABLs were run for G-B channels, and Spearman correlation analysis were for R channel because the distributions of MABLs and the investigation parameters in R channel are not normal (The normality was examined by Shapiro-Wilk test sig>0.05 after removing some outliers). The both of the correlation coefficient range are (−1, 1), where the Pearson coefficient (PCC) close to 1 or -1 indicates a perfect linear relationship and the value of that close to 0 demonstrates no relations, while the Spearman coefficient of +1 or −1 indicates a perfect monotonic relationships between two variables. Table I shows the correlation coefficients.

TABLE I: PEARSON CORRELATION COEFFICIENTS (PCC) AND SPEARMAN COEFFICIENTS (SPC) BETWEEN MABLS AND THE AVG, MED, MAX, MIN AND STD IN CORRESPONDING TO EACH R-G-B CHANNEL.

| Channel | Avg | Med | Max | Min | Std |
|---|---|---|---|---|---|
| R | 0.824** | 0.844** | 0.269** | 0.639** | 0.216** |
| G | 0.687** | 0.672** | 0.017** | 0.264** | 0.186** |
| B | 0.742** | 0.719** | 0.274** | 0.149** | 0.405** |

One can readily see that there is a significantly strong correlation between MABLs and average value or median value, following by the standard deviation. Although strong correlations can also be seen at Max and Min for R channel, they are redundant parameters due to the strong correlations between Min and Med (0.677) and between Max and Std (0.783). A strong correlation also exist between Avg and Med. Therefore, we only choose the Avg/Med and Std as the main predictors for the annotated BLs.

According to the relation of the selected parameters and the MABLs, firstly we define a linear model of the Avg and the Std for the BL estimation of the G-B channels as follows:

$$B^{c'} = \alpha \times Avg^{c'} + \beta \times Std^{c'} + \gamma \quad (15)$$

where $Avg^{c'}$ and $Std^{c'}$ are the average and the standard deviation of channel $c'$ of the input image respectively; $\alpha$ and $\beta$ are coefficients; $\gamma$ is a constant. The linear regression model for capturing correlation between MABLs of GB channels and selected two parameters is selected because it is simple but sufficient for our purposes.

As for R channel, a non-linear model is defined in (16) based on curve estimation.

$$B^r = \frac{a}{1+b \times Exp(c \times Med_r)} \quad (16)$$

where $a, b, c$ are the coefficients. To avoid the effect of noise or extreme values of pixels, we used the middle 80% of the entire channel intensity histogram of coarse underwater images (limited to 10% from the lower and upper parts) to calculate the Avg, Med and Std. Through linear regression and non-linear regression under 10-fold cross validations, we eventually determined these coefficients in Eq. (17)-(18) as below.

$$B^{c'} = 1.13 \times Avg^{c'} + 1.11 \times Std^{c'} - 25.6 \quad (17)$$

$$B^r = \frac{140}{1+14.4 \times Exp(-0.034 \times Med_r)} \quad (18)$$

For avoiding producing over-fitting or under-fitting due to the limited MABLs, we will empirically restrict the value of the estimated BLs by the above the simple model of the BLs estimation between 5 to 250, so the final estimated BL is as follows:

$$B^c = min(max(B^{c',r}, 5), 250), c \in \{r, g, b\} \quad (19)$$

For the linear model, the Adjusted $R^2$ is above 0.6. For non-linear model, $R^2$ is bigger than 0.65, computed by: 1- the residual sum of squares/ the corrected total. The R2 value means that the models account for about 60% of the variability in BLs. Pearson correlation coefficient (PCC), Spearman coefficient (SPC), the $R^2$ and adjusted $R^2$ are all statistical analysis terms. The investigation parameters in the GB channels have a strongly linear relation to the distributions of MABLs, thus we used PCC to indicate their correlation (PCC is a measure of the linear correlation between two variables X and Y). But the investigation parameters in R channel are not linear with MABLs, so the SPC that can assess monotonic relation (whether linear or not), is used to measure their correlation. Values in table I are obtained according to the PCC and SPC between MABLs and the Avg, Med ... in corresponding to each R-G-B channel of 350 underwater images. The $R^2$ and adjusted $R^2$ are squared $R$ to justify the performance of linear regression model. They give us an idea of how many data points fall within the line of the regression equation, which can account for the majority of the variability in BLs.

### B. TM Estimation Model

Due to the similarities between the outdoor haze images and the underwater images, the TMs were estimated using DCP [15], [16] and its variation UDCP [18], which were under the assumptions of $J^{rgb}_{dark} = 0$ and $J^{gb}_{dark}(x) = 0$, respectively. Even though the DCP and UDCP assumptions seem to be proper, using them in the underwater images would arise some problems because they both ignored the wavelength independence in the water medium.

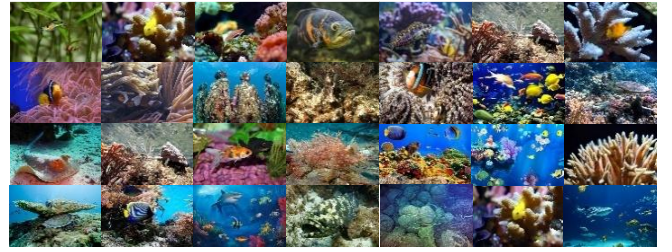

Fig. 5: Samples of our high-quality underwater images

Inspired by the observation of He et al. [15] that a clear outdoor image is approximately zero in the local patch at least one of the RGB color channels, we propose a new underwater dark channel prior by the statistics of high-quality underwater images. Following the guidelines of He's outdoor image selection from several types of haze-free images, we selected some clear and high resolution (not less than $1280 \times 720$ pixels) underwater images from Google Images and shutterstock.com, including underwater animals, various marine scenes, coral reefs, rocks, archaeological ruins and divers. Some clear underwater images are shown in Fig. 5. In order to verify our prior of underwater images, all selected images are resized to the size of $400 \times 600$ pixels and their dark channels are computed using the local patch size of $13 \times 13$ pixels.

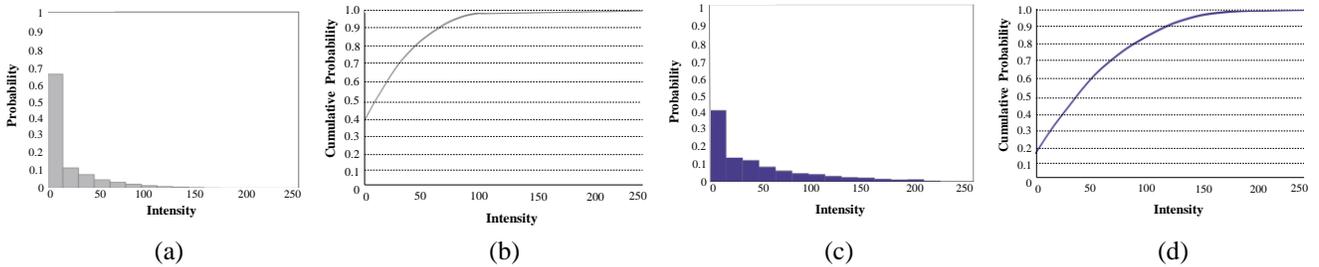

(a) (b) (c) (d)

Fig. 6: Statistics of the R-G-B dark channel and G-B dark channel based on our high-quality underwater images database. (a) The histogram distribution of the R-G-B dark channel in all of 1000 images (each bin stands for 16 intensity levels), (b) Cumulative distribution corresponding to fig. (a), (c) The histogram distribution of the G-B dark channel in all of 2000 images (each bin stands for 16 intensity levels), (d) Cumulative distribution corresponding to fig. (c).

In order to augment the limited image database, each image is downscaled by factors of 0.5 to 1.0 by 0.1, up-scaled by factors of 1.1 to 1.5 by 0.1, and then flipped vertically and horizontally. 2000 free-hazed underwater images are used for computing the histograms and cumulative distributions of pixel values. Fig. 6 shows the statistics results of the R-G-B dark channel (NUDCP, called as new underwater $DCP_{rgb}$) and G-B dark channel (UDCP, called as $DCP_{gb}$) based on our high-quality underwater images database in terms of the histogram distributions and cumulative distributions. Each bin in Fig. 6 stands for 16 pixel intensities in an interval of (0, 255).

Fig. 6 (a-b) represents the histogram distributions of the R-G-B dark channel and the corresponding cumulative distribution. Fig. 6 (c-d) is similar with Fig. 6 (a-b) but represents distributions of G-B dark channel. Although the distributions of Fig. 6 (a-b) show some similarities of the distribution shape with that of DCP [10], the difference is much less ratios of zero values in our distributions (see Fig. 6 (a-b)) than those in He's distribution. According to the distributions of Fig. 6 (a-b), the statistical results based on R-G-B channels using our high-quality underwater images show the probabilities of 0, (1, 15), (16, 47) are approximately 40%, 20% and 20% respectively, while He's statistic showed about 75% of the pixels were 0, and 90% were below 25. From the distributions of Fig. 6 (c-d), we can observe an even lower probability of zero values, which violates the assumption of DCP proposed by He. Therefore, we argue that it is inappropriate to simply set $J_{dark}^{rgb}(x)$ or $J_{dark}^{gb}(x)$ as 0 for underwater images. We determine a new underwater dark channel prior according to the pixel values of the probability distribution in the Fig. 6 (a-b), which suggests the value of underwater dark channel prior should be set as 25. After normalization, we set $J_{dark}^{rgb} = 0.1$ as the result of new underwater DCP.

Firstly, we take the minimization operation in the local patch $\Omega$ on the hazy image in Eq. (1):

$$\min_{y \in \Omega}(I^c(y)) = \min_{y \in \Omega}\{J^c(y)t^c(y)\} + \min_{y \in \Omega}\{(1 - t^c(y))B^c\} \quad (20)$$

The BL is the homogeneous background light and an accurate BL is estimated (the minimum value is greater than 0), so both sides of Eq. (20) can be divided by $B^c$:

$$\frac{\min_{y \in \Omega}(I^c(y))}{B^c} = \frac{\min_{y \in \Omega}\{J^c(y)t^c(y)\}}{B^c} + \min_{y \in \Omega}\{1 - t^c(y)\} \quad (21)$$

The TM is essentially constant on the small local patch, the Eq. (21) can be described as:

$$\frac{\min_{y \in \Omega}(I^c(y))}{B^c} = \frac{\min_{y \in \Omega}(J^c(y))}{B^c} t^c(x) + (1 - t^c(x)) \quad (22)$$

The final minimum filter is performed among three color channels as follows:

$$\min_c \left\{\frac{\min_{y \in \Omega}(I^c(y))}{B^c}\right\} = \min_c \left\{\frac{\min_{y \in \Omega}(J^c(y))}{B^c} t^c(x)\right\} + 1 - \min_c\{t^c(x)\} \quad (23)$$

The first term on the right-hand side of Eq. (23) can be express as the following equality:

$$\min_c \left\{\frac{\min_{y \in \Omega}(J^c(y))}{B^c} t^c(x)\right\} = \min_c \left\{\frac{\min_{y \in \Omega}(J^c(y))}{B^c}\right\} \times \min_c\{t^c(x)\} \quad (24)$$

The first term on the right-hand side of Eq. (24), labelled as $V$, can be express as the following inequality:

$$\frac{\min_c\{\min_{y \in \Omega}(J^c(y))\}}{\max_c\{B^c\}} \leq \min_c \left\{\frac{\min_{y \in \Omega}(J^c(y))}{B^c}\right\} = V \leq \frac{\min_c\{\min_{y \in \Omega}(J^c(y))\}}{\min_c\{B^c\}} \quad (25)$$

Because of $\min_c\{\min_{y \in \Omega}(J^c(y))\} = 0.1$ based on our new DCP, the following inequality is shown as:

$$\frac{0.1}{\max_c\{B^c\}} \leq V \leq \frac{0.1}{\min_c\{B^c\}} \quad (26)$$

According to $t_\lambda(x) = Nrer(\lambda)^{d(x)}$ in Eq. (2), among RGB channels, $Nrer(Red)$ is the lowest residual value in the same local patch. Hence $\min_c\{t^c(x)\}$, an be expressed as $t^r(x)$, therefore, Eq. (23) can be rewritten as:

$$\min_c \left\{\frac{\min_{y \in \Omega}(I^c(y))}{B^c}\right\} = V \times t^r(x) + 1 - t^r(x) \quad (27)$$

And $t^r(x)$ can be obtained as:

$$t^r(x) = \left(1 - \min_c \left\{\frac{\min_{y \in \Omega}(I^c(y))}{B^c}\right\}\right)/(1 - V) \quad (28)$$

Meanwhile, the $t^r(x)$ ranges from zero to one. When the bigger $V$ is, the $t^r(x)$ is more possibly to be greater than 1. In order to minimize the information loss of the TM, in this paper, $V$ is set as $\frac{0.1}{\max_c\{B_c\}}$ and $\max_c\{B_c\}$ is simply set as $B_{max}$. Finally, Eq. (26) can be rewritten as:

$$t^r(x) = \frac{1 - \min_c\left\{\frac{\min_{y \in \Omega}(I^c(y))}{B^c}\right\}}{1 - 0.1/B_{max}} \quad (29)$$

To use the suitable transmission map $t^r(x)$ in the IFM, $t^r(x)$ is stretched to a proper range ($O_{min}$, $O_{max}$) following the histogram stretching [38] as:

$$p_{out} = (p_{in} - I_{min})\left(\frac{O_{max} - O_{min}}{I_{max} - I_{min}}\right) + O_{min} \quad (30)$$

where $p_{in}$ and $p_{out}$ are the input and output pixels, respectively, and $I_{min}$, $I_{max}$, $O_{min}$ and $O_{max}$ are parameters for the before and after stretched transmission map, respectively. To reduce the under- and over-stretched effects, 0.2% from the lower and upper range of the output histograms are stretched to the desired range of the red transmission map is set as (0.1, 0.9).

Some underwater images and corresponding coarse TMs of red channel are shown in the Fig. 7. According to the Fig. 7 (a)-(c), our proposed NUDCP can successfully estimate TMs of these standard underwater images which are constituted by the dim foreground scene and the light background scene. In the Fig. 7 (d), the white fish in the front of the image is regarded as the background scene, and our proposed method failed to identify the distance from the camera to a near fish and the far ground. It can be seen in the Fig. 7 (e) that some part region of TM seem overestimated due to the low intensity of the red channel but relatively high intensity compared with the BL of red channel. Fig. 7 (f) displays an artificially illuminated scene and an incorrect TM, which is calculated by our proposed assumption and understands the foreground region with AL farther to the camera. These problems which are introduced by the DCP, new UDCP and variations of DCP, can be simply solved by the following two steps, including the compensation by the depth map and the optimization by the ARSM.

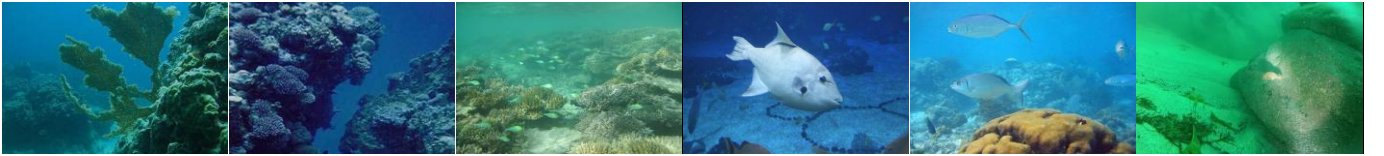

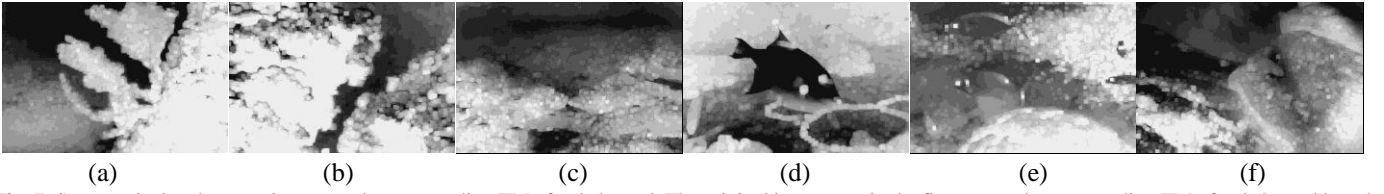

(a) (b) (c) (d) (e) (f)

Fig. 7: Some typical underwater images and corresponding TM of red channel. The original images are in the first row and corresponding TM of red channel based on the assumption of NUDCP. (a)-(c) in character of the dim foreground and light background obtain accurate TMs, but the (d)-(f) containing with white fish, the relatively dim component of red channel, and the artificial light shot in the foreground stone estimate apparently wrong TMs.

In our previous work [A Rapid Scene Depth Estimation Model Based on Underwater Light Attenuation Prior for Underwater Image Restoration, Reference], we reveal underwater light attenuation prior (ULAP) that the scene depth increases with the higher value of the difference between the maximum value of G and B lights and the value of the R light. In the other word, the difference between the maximum value of G-B intensity (simplified as MVGB) and the value of R intensity (simplified as VR) in one pixel of the underwater image is very strongly related to the change of the scene depth. Based on the ULAP, we define a linear model of the MVGB and VR for the depth map estimation as follows:

$$d(x) = \mu_0 + \mu_1 m(x) + \mu_2 v(x) \quad (31)$$

where $x$ represents a pixel, $d(x)$ is the underwater scene depth at point $x$, $m(x)$ is the MVGB, $v(x)$ is the VR. The best learning result is $\mu_0 = 0.53214829$, $\mu_1 = 0.51309827$ and $\mu_2 = -0.91066194$. In order to operate edge-preserving smoothing on the coarse estimated depth maps, the guided filter [Guided Image Filtering, Reference] is used to refine the coarse depth maps.

The estimated depth maps are the relative distance in the image. To measure the absolute distance from the camera to each scene point, the distance $d_0$ between the closest scene point and the camera must be estimated in advance. Via the maximum difference between the estimated $B^c$ and the input image $I^c(x)$, the base depth $d_0$ can be calculated by:

$$d_0 = 1 - \max_{x,c}\left(\frac{|B^c - I^c(x)|}{\overline{B^c}}\right) \quad (32)$$

where $\overline{B^c} = \max(1 - B^c, B^c)$, $d_0 \epsilon [0,1]$. The denominator in the Eq. (32) is used as the normalization processing. The numerator in the Eq. (32) is the absolute difference between the observed intensity and the background light, and the point within the larger value is closer to the camera. Hence the actual scene depth map $d_a$ is defined as follows:

$$d_a(x) = D_\infty \times (d(x) + d_0) \quad (33)$$

where $D_\infty$ is a scaling constant for transforming the relative distance to the real distance, and in this paper, the $D_\infty$ is set as 10.

With the estimated $d_a(x)$, we can calculate the TMs for R-G-B channels based on the Eq. (2), select the $Nrer(\lambda)$ for R-G-B light as 0.83, 0.95 and 0.97, respectively, and set TM of the red channel as $t_r^{ULAP}(x)$. In the Fig. 7 (e), due to the small values of the background light and the unreliable observed intensity in the red channel, the TM estimation is controlled by the red channel and the background region failed to be estimated as the farther distance. The ULAP mainly considers that the bigger difference between the value of GB channel and the value of R channel is the farther depth, which can be used to compensate and modify the TM estimation based on the NUDCP and the enhanced estimation of $t^r(x)$ is computed as:

$$t_{cps}^r(x) = \sum_{i=1}^{M}\sum_{j=1}^{N} min\big(t^r(i,j), t_{ULAP}^r(i,j)\big) \quad (34)$$

where $M, N$ denotes the width and height of the image, respectively through this paper. The minimum filter is used to adjust the overestimated TM due to the low intensity of the red channel when using the NUDCP.

So far, we have completed compensated the impact of the red channel in estimating the TM, but we haven't taken the influence of the artificial light (AL) into consideration. We found that the region with AL in the Fig. 7 (f) (the patch in that region has strong intensities of red channel in the green-bluish underwater image) has low saturation in HSV color model. Saturation defined in Eq. 36, describes the purity of the chromaticity of a pixel.

$$Sat(I^c(x)) = \sum_{i=1}^{M}\sum_{j=1}^{N} \frac{max(I^c(x)) - min(I^c(x))}{max(I^c(x))}$$

$$Sat(I^c(x)) = 1, if\ max(I^c(x)) = 0 \quad (36)$$

When a color is fully saturated, it is considered in purest (truest) version, that is to say, a color will lose saturation when we add white light which contains power at all wavelengths. Hence the region with lack of saturation in the image can be interpreted as a large amount of white light shooting in that region. Especially for underwater images, the saturation of the scene without AL is notably greater than that under artificially illuminated areas. We can express this phenomenon by a reversed saturation map (RMS) defined as Eq. 37. The region with high values of RSM often means the AL affected region.

$$Sat^{rev}(x) = 1 - Sat(I^c(x)) \quad (37)$$

Fortunately, because the values in artificially illuminated areas is greatly higher than values in the other parts of the image, almost trend to zero, the areas with the AL in underwater images are often close to the camera which can be represented as the correct TM by the Eq. (37). Then, we can use the RSM to optimize the estimated TM to reduce the effect of the artificial light. To modify relative values of the TM based on the reversed saturation, $\lambda \epsilon [0,1]$ is imported to the RSM as an effective scalar multiplier.

$$Sat_{adj}^{rev}(x) = \lambda * Sat^{rev}(x) \quad (37)$$

The RSM and adjusted RSM (ARSM) of the Fig. 7(f) are shown in Fig. 8. It can be seen that the areas under the scene of AL are segmented effectively from the rest of the image. Fig. 8 (b) shows the ARSM when the $\lambda$ is set as 0.7. In the optimization of modified TM, we can change the values of the areas under artificial light using different $\lambda$.

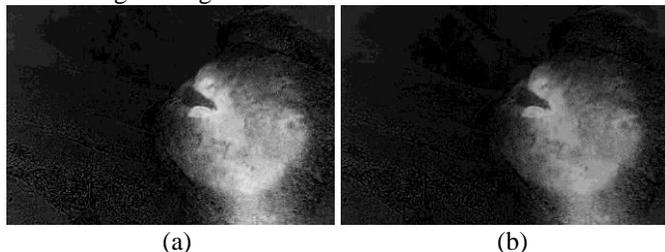

(a)                  (b)

Fig. 8: The RSM and ARSM of the original of Fig. 7. (f) without and with effective scalar multiplier $\lambda$. (a) RSM obtained by Eq. (36), (b) ARSM obtained by Eq. (37) when $\lambda = 0.7$.

When the ARSM is estimated, the transmission information in the artificial illuminated areas can be simply applied to modify the error TM, preserving the transmission information of the other regions due to the low-intensity regions in ARSM without artificial light. Hence the final optimal TM of the red channel can be expressed as:

$$t_f^r(x) = \sum_{i=1}^{M} \sum_{j=1}^{N} max\left(t_{cps}^r(i,j), Sat_{adj}^{rev}(i,j)\right) \quad (38)$$

where the maximum filter is used to extract the high-layer transmission information and retain the low-layer transmission information. When the $\lambda$ is set as zero, the maximum filter in the Eq. (38) becomes useless and the final TM of red channel is obtained by Eq. (34).

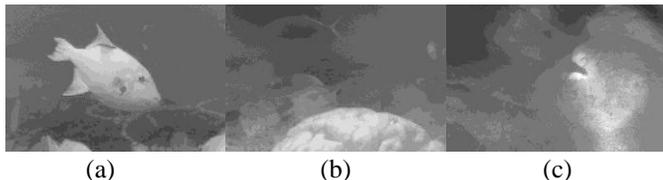

(a)             (b)             (c)

Fig. 9: Optimal TMs of R channel in the Fig. 7 (d)-(f). (a) the white fish in the foreground scene is correctly estimated as larger values closer to the camera, (b) the values of TM in background scene is compensated by the depth map obtained based on the ULAP, (c) the TMs for the underwater image spotted by the AL have larger values for the scene points closer the light and smaller values for the points farther from the light.

By means of a series of the compensated and modified process on the new UDCP-based TM, the inaccurate TMs in the Fig. 7 (d)-(f) can be rectified into correct TMs, shown in Fig. 9, which are refined by the guided filter [39] and stretched by the histogram stretching. Basis on the Eq. (2), the scene depth map is inversed as

$$d(x) = log_{Nrer(Red)}\{t_f^r(x)\} \quad (39)$$

The TMs for blue and green channels are computed based on the exponential relationship between normalized residual energy ratio $Nrer(c')$ and the depth of object to the camera $d(x)$, and then deducted as Eq. (40):

$$t^{c'}(x) = Nrer(c')^{d(x)} \quad (40)$$

Because the coarse estimated TM are calculated over a local patch of the underwater image, some halos and block artifacts are produced in the transmission map $t^c(x)$. In order to solve these problems, the guided filter [39] is further used to refine the coarse TMs.

Lastly, the recovered underwater image $J^c$ is obtained by applying refined TM and an accurate BL of RGB channels into the restored equation (41):

$$J^c = \frac{I^c(x) - B^c}{\min(\max(t^c(x), 0.2), 0.9)} + B^c, c \in \{r, g, b\} \quad (41)$$

where $J^c$ is the restored underwater image, a lower bound and an upper bound for $t^c(x)$ empirically set to 0.2 and 0.9, respectively. Fig. 10 (a) shows the typical underwater image with the artificial light [Underwater Image Enhancement by Wavelength Compensation and Dehazing, Reference], and Fig. 10 (b)-(d) gives an example of TMs for red channel based NUDCP, the compensation of the depth map and the optimization of ARSM. Fig. 10 (e) show the restored result.

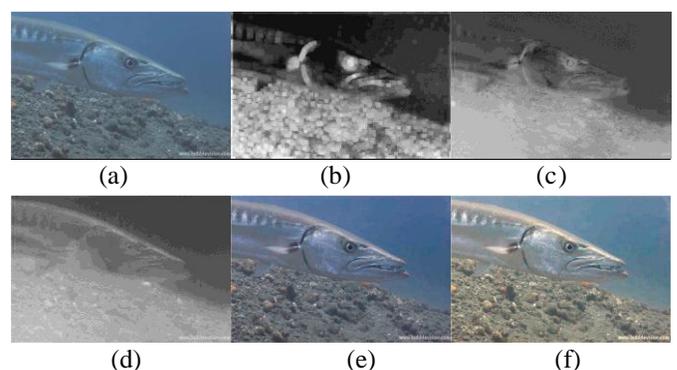

(a)             (b)             (c)

(d)             (e)             (f)

Fig. 10. The entire processing of underwater image enhancement. (a) Original underwater image with artificial light under the depth from 17 to 18 [17], (b) Coarse TM of red channel obtained by NUDCP, (c) Refined TM of red channel compensated by TM of red channel estimated by ULAP, (d) Refined TM of red channel modified by the reversed saturation, (e) Restored image, (f) Enhanced image improved by color correction.

### C. Color Correction

Although the underwater image can be dehazed by using restoration parameters of the TMs and BLs, the restored image is often characterized by low brightness and contrast, which veils many valuable image details. The proposed color correction is based on the white balance algorithm with optimal gain factor [26], [40] and can be described as:

$$\begin{cases} P_o = \frac{P_i}{\lambda_{max} \times (\mu/\mu_{ref}) + \lambda_v} \\ \mu_{ref} = \sqrt{(\mu_r)^2 + (\mu_g)^2 + (\mu_b)^2} \end{cases} \quad (42)$$

where $P_o$ and $P_i$ denote the color corrected image and the original coarse underwater image. $\mu = \{\mu_r, \mu_g, \mu_b\}$ represents the average value of each R-G-B channel of the input underwater image $P_i$, and $\lambda_{max}$ is estimated by the maximum value of R-G-B channels of the input underwater image $P_i$. The value of $\lambda_v$ is selected in the range of (0, 0.5) to get the desired color for the enhanced image. The closer $\lambda_v$ is to 0, the lower the brightness of the corrected image is.

Fig. 11 displays some underwater images enhanced based on different values of $\lambda_v$. According to plenty of experimental results, when the value of $\lambda_v$ is 0.2, the corrected image is slightly bright, and when $\lambda_v$ is 0.3 the corrected image is a little dim, thus the optimal $\lambda_v$ is chosen as 0.25 in this paper. Fig. 10 (f) show the enhanced result, which preserves the actual color tone and improves visibility and contrast of restored image.

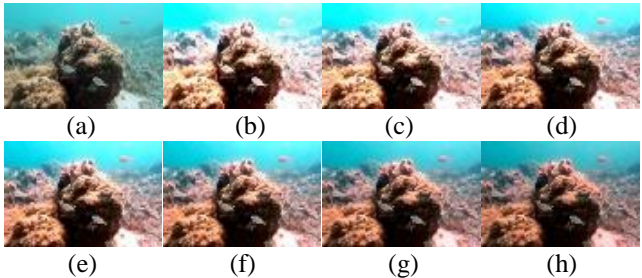

Fig. 11. Results of color correction based on different values of $\lambda_v$. (a) Original image, (b)-(h) are the enhanced images with $\lambda_v$ 0, 0.1, 0.2, 0.25, 0.3, 0.4 and 0.5 respectively.

## IV. RESULTS AND EVALUATION

In this section, in order to ensure the fairness of each evaluation system, all test underwater images are pre-processed as the size of 400×600 pixels. All methods are implemented on a Windows 7 PC with Intel(R) Core(TM) i7-4790U CPU@3.60GHz, 8.00GB 1600MHz DDR3 Memory, running Python3.6.3.

### A. Evaluation of Objectives and Approaches

We have conducted experiments on four objectives:
a) To examine the effectiveness of the proposed MABLs;
b) To assess the performance of the statistical BL model;
c) To assess the performance of the TM optimizer;
d) To assess the overall performance of the proposed underwater image enhancement method.

The evaluations were achieved by comparing with underwater image restoration/enhancement methods and using five quality assessment metrics. These methods include Maximum Intensity Prior (MIP) method [24], Dark Channel Prior (DCP) method [15], [16], underwater image enhancement with a new optical model [37], Underwater Dark Channel Prior (UDCP) method [18] and extended in [19], depth estimation based on blurriness [22], Red channel method [20], Li's method [21] and Peng's method based on image blurriness and light absorption [23]. Limited by the space, we cannot compare our method with all methods, hence the typical methods, including DCP [16], MIP [24], UDCP [18], Li [21], Peng [23] are selected to demonstrate the superiority of our method in terms of the BLs and TMs estimation.

In order to objectively evaluate the performance of various underwater image restoration/enhancement methods from different aspects, we employed five image quality metrics and the running time (RT). The full-reference quantitative metrics are root mean square error (RMSE) and the structural similarity (SSIM) [41]. The RMSE mainly measures the degree of noise in the image, and a smaller RMSE indicates better performance and vice versa. The SSIM is the most prominent approach introduced to evaluate the ability to preserve the structural information of the images. A higher SSIM represents high similarity between the dehazed image and the ground truth image and vice versa. RMSE and SSIM require the original underwater image as the reference image, they are still useful to indicate the introduced artificial noise of the enhanced image and the retention of structural information of the original image. Meanwhile, we adopt three non-reference quantitative metrics: Entropy (S), the Blind/Referenceless Image Spatial Quality Evaluator (BRISQUE) [42], and Underwater color image-quality evaluation (UCIQE) [43]. Entropy represents the abundance of information. A higher entropy value of an image states more valuable information contained in the image. The BRISQUE quantifies possible losses of naturalness in an image due to the presence of distortions. The BRISQUE value indicates the image quality from 0 (best) to 100 (worst). The UCIQE in a linear combination of chroma, saturation, and contrast evaluates the degree of the non-uniform color cast, blurring and low-contrast in the underwater image. It is in a range of (0, 1), and the higher the better.

### B. Effectiveness of Proposed MABLs

According to the principle of choosing the far scene points with high intensity and the biggest difference of R channel and G-B channels, a manually annotated background lights (MABLs) database has been established with 500 underwater images. But a question is: are the MABLs proper?

To answer this question, we conducted a series of experiments using the MABLs and the BLs generated by the methods of DCP, MIP, UDCP, Li's method, Peng's method and our proposed method to restore the 500 underwater images, given the same TMs. Then, we evaluated the restored image qualities under different BLs. Our assumption was: the highest quality would be seen on the images restored with the MABLs. To avoid the influence of TMs, we adopted and improved the TMs from Peng's method [23] based on the light absorption and image blurriness for all the experiments. Some transmission maps exist obvious estimation errors (e.g., a close-up fish is black in the map) were discarded, and 100 fully-proper transmission maps are well-chosen by the final manual selection. Table II gives the quality assessment results. It can be seen for all the five quality assessment metrics, MABLs achieves the best performance. Fig. 12 show different kinds of underwater image examples restored by using MABLs. Besides Peng's TMs, we also conducted the experiments under other TMs proposed in MIP, UDCP and Li's method. The results indicated the MABLs always performed the best, though the final results had different brightness or color.

It should be noticed that we cannot guarantee the absolute correctness of MABLs. For some images with a close shot, the far scene point is difficult to identify. Thus, the MABLs may not work out a significant improvement of quality, but they at least do not cause distortion. In fact, for that kind of underwater images, none of the BL estimation methods (experimented in this paper) could do a good job.

TABLE II: Quantitative analysis via RMSE, SSIM, Entropy, BRISQUE, UCIQE and RT of restoration results based on different BLs and all the TMs [23].

| Methods | Indexes | | | | | |
|---|---|---|---|---|---|---|
| | RMSE | SSIM | S | BRISQUE | UCIQE | RT (s) |
| DCP | 39.12 | 0.52 | 6.72 | 34.05 | 0.32 | 0.92 |
| MIP | 36.38 | 0.53 | 6.64 | 33.14 | 0.36 | 1.97 |
| UDCP | 35.95 | 0.55 | 6.71 | 33.62 | 0.38 | 0.73 |
| Li | 28.57 | 0.76 | 7.36 | 29.55 | 0.46 | 1.48 |
| Peng | 31.62 | 0.71 | 6.98 | 30.37 | 0.41 | 1.55 |
| Ours | 23.64 | 0.81 | 7.67 | 28.56 | 0.49 | **0.06** |
| MABLs | **21.99** | **0.83** | **7.88** | **28.15** | **0.53** | - |

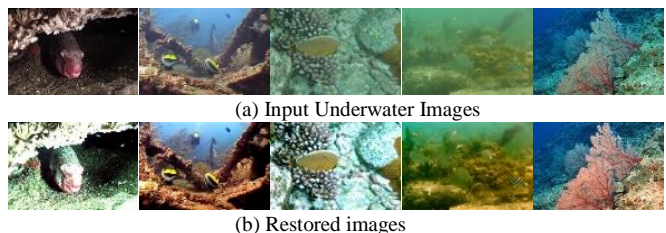

(a) Input Underwater Images

(b) Restored images

Fig. 12: Successful results based on the MABLs with the TMs [23].

### C. Statistical Model of BL Estimation

#### 1) Discussion on BL accuracy and efficiency

We applied the established statistical models to estimate the BLs of 500 images in MABLs database. Then, we computed the differences between the predicted BLs and the MABLs, using 30% testing data. Finally, we labeled whether each predicted BL was accurate with a tolerant range of 30 for R channel and 40 for G-B channels. The tolerance was set because MABLs do have uncertainty to some extend and a small deviation of BL does not have a great impact on the quality of restored images based on our experiments. The same labelling work has done on the BLs predicted by DCP, MIP, UDCP, Li's and Peng's methods. The compared results of prediction accuracy and the running time are shown in Fig. 13.

From the Fig. 13 (a), the accuracy of BLs estimated by DCP and UDCP are two lowest of all methods, further to show that the DCP and UDCP cannot be suitable for different kinds of underwater images. The overall prediction accuracy of the other three compared methods is obviously lower than our proposed statistical model. Due to the consumption of vast computation in the dark channel of G-B channels or R-G-B channels, the running time of DCP, UDCP, Peng' method and Li' method are in linear increases as the sizes (/pixels) of test underwater images are larger. The running time of MIP is the longest with different sizes of underwater images because the maximum intensity is used twice (maximum intensity of R channel and G-B channels). A non-linear model for R channel and linear model for G-B channels only need limited time to determine the BLs of RGB channels when estimating the effective BLs with the best accuracy.

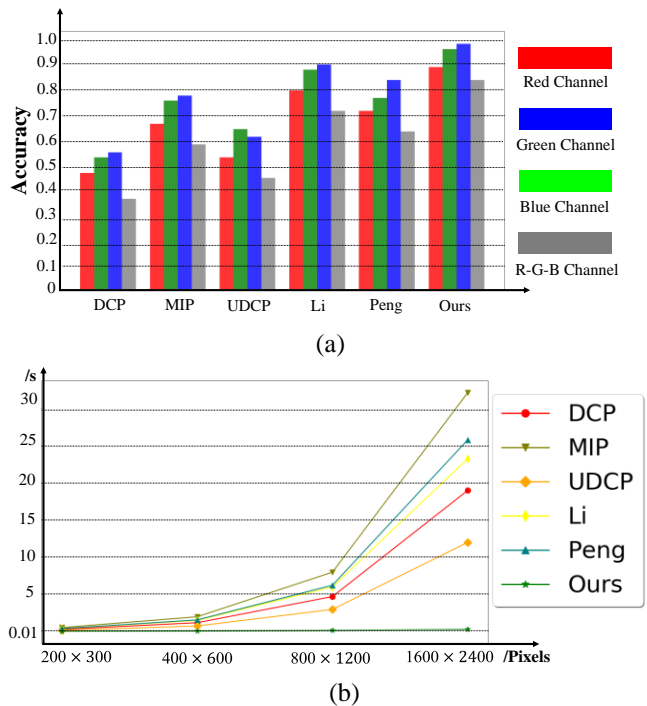

Fig. 13: (a) Accuracy of the estimated BLs in the reference with MABLs. (b) The running times (/s) with different sizes (/pixels) of test underwater images.

The overall performance of our statistical model of BLs estimation can be seen in Table 3. Except the MABLs, our method outperforms other five methods in terms of image quality assessment. The running time (RT) is also significantly lower than others. Our BL estimation could save 1000% of time than the fastest UDCP method.

#### 2) Restoration of Images based on the Statistical Model

Fig. 14 demonstrates the restored images captured in sunlight scene and diving scene, where the same TMs computed by improved Peng's method. The DCP and UDCP methods for estimating BLs in Fig. 14 (b) and (d) fail to improve the contrast and color, even cause serve distortion, which explains that these methods cannot obtain the correct BLs of underwater images under these complicated scenes. The method proposed by Li (Fig. 14 (e)) can produce nearly similar satisfactory results with the MIP method (Fig. 14 (c)) due to BLs estimation of all methods based on the maximum difference between R channel and G-B channels. This is also in our principle of creating the MABLs. But for the diver image under deep water, the restored image by Li and Peng becomes relatively dimmer than the original due to estimating the bigger BLs. Our proposed BLs estimation can successfully improve the color and contrast and the restored images just appear different colors.

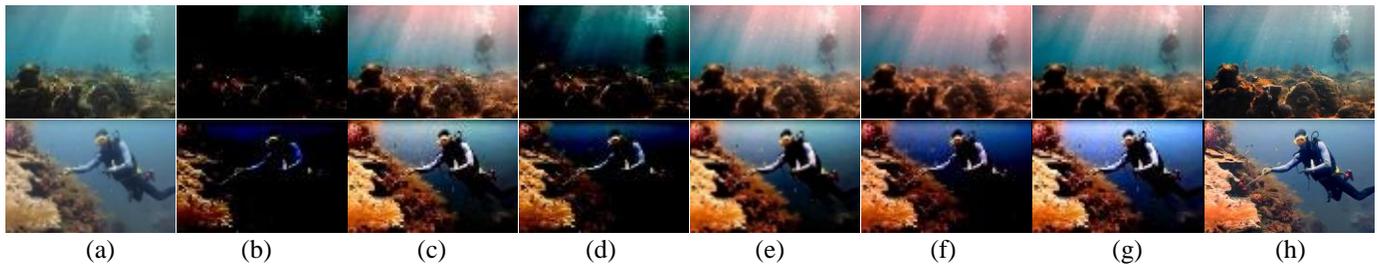

Fig. 14: Comparisons results with different color tones based on different BLs and the same TM. (a) Input images. (b)-(i) Restored results with different BLs proposed by (b) DCP, (c) MIP, (d) UDCP, (e) Li, (f) Peng, (g) Ours, (h) MABLs.

### D. Performance of TM Optimizer

Since the above experimental results have adequately demonstrated that the MABLs are best suitable to diverse underwater image restoration, we examined the performance of TM estimations based on the MABLs. Fig. 15 shows the restored images using different TMs.

On account of DCP, MIP and UDCP methods generating the only TM for all R, G and B channels, they ignore the crucial difference between TM of R channel and the TMs of G-B channels. As a result, the restored coral and fish images exist some over-saturation areas, seen in Fig. 15 (b-d). From the Fig. 15 (e) and (f), the TMs obtained by Peng's and our method can increase the image valuable information, entire contrast and local details. To conclude, the R and G-B TMs should be deducted from the characteristics of the corresponding channel separately.

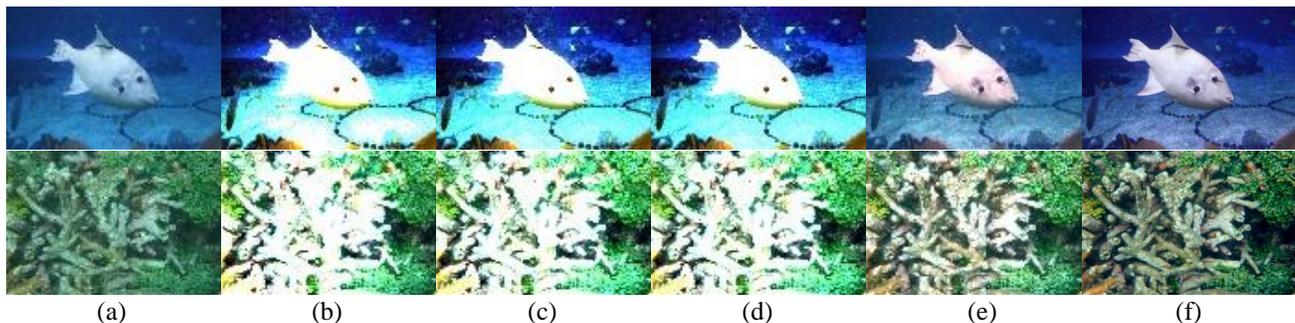

Fig. 15: Comparison results restored with different TMs based on the MABLs. (a) Original images. The restored results by (b) DCP, (c) MIP, (d) UDCP, (e) Peng, (f) Ours.

From the full-reference and non-reference quality analysis in Table III, the results of DCP, MIP and UDCP methods are extraordinary lower than the results of our method and Peng's. Although the results of Peng's method have a little better than the results of our method in terms of SSIM and BRISQUE, the running time of Peng is nearly double the time of our proposed method.

TABLE. III: Quantitative analysis via RMSE, SSIM, Entropy, BRISQUE, UCIQE and RT of restoration results based on different estimated TMs and MABLs (the best results in bold)

| Methods | Indexes | | | | | |
|---|---|---|---|---|---|---|
| | RMSE | SSIM | S | BRISQUE | UCIQE | RT (s) |
| DCP | 35.95 | 0.50 | 6.83 | 33.91 | 0.37 | 0.63 |
| MIP | 34.76 | 0.53 | 6.86 | 32.45 | 0.39 | 1.38 |
| UDCP | 30.11 | 0.51 | 6.98 | 31.71 | 0.42 | 0.59 |
| Li | 26.18 | 0.79 | 7.83 | 29.13 | 0.49 | 0.93 |
| Peng | 23.64 | **0.82** | 7.97 | **28.27** | 0.54 | 1.24 |
| Ours | **22.31** | 0.81 | **7.99** | 28.56 | **0.53** | **0.67** |

### E. Overall Performance of the Proposed Method

The objective of the proposed our proposed underwater image enhancement method is not only to recover the underlying scene radiance but also to enhance the contrast of underwater image, preserve the genuine color and improve the visibility of the input images. Therefore, a color correction with white balance was added after restoration. To fairly evaluate the overall performance of our method, we compared our method with the IFM based methods DCP, MIP, UDCP, Peng, as well as these methods with a post-processing of histogram equalization (HE). We also compared with Li's method [21], which introduces the red channel correction and adaptive exposure map estimation as color correction. For convenience, our proposed method and ours without contrast correction are represented by Ours and OWCC.

TABLE IV: Quantitative analysis via RMSE, SSIM, Entropy, BRISQUE, UCIQE and RT of restoration and enhancement results based on different methods

| Methods | Indexes | | | | | |
|---|---|---|---|---|---|---|
| | RMSE | SSIM | S | BRISQUE | UCIQE | RT (s) |
| DCP | 45.12 | 0.48 | 5.57 | 48.05 | 0.29 | 1.46 |
| MIP | 43.95 | 0.49 | 6.21 | 46.28 | 0.32 | 2.58 |
| UDCP | 47.56 | 0.56 | 5.81 | 41.18 | 0.34 | 1.02 |
| Peng | 30.59 | 0.71 | 7.38 | 35.78 | 0.52 | 2.21 |
| OWCC | 30.45 | 0.72 | 7.66 | 34.59 | 0.54 | **0.69** |
| Li | 32.45 | 0.68 | 7.42 | 35.73 | 0.52 | 1.86 |
| DCP+HE | 37.59 | 0.51 | 6.64 | 44.12 | 0.43 | 1.52 |
| MIP+HE | 38.82 | 0.58 | 6.81 | 43.88 | 0.46 | 1.61 |
| UDCP+HE | 39.28 | 0.58 | 6.71 | 37.78 | 0.47 | 1.13 |
| Peng+HE | 28.81 | 0.73 | 7.74 | 33.58 | 0.58 | 2.33 |
| Ours | **27.45** | **0.75** | **8.03** | **31.78** | **0.63** | 0.71 |

The quantitative assessments, shown in Table IV, present the average values of 150 test images outside our MBALs database. Fig. 16 shows two examples of the comparison results. From Table IV, we can see that adding HE to the methods of DCP, MIP, UDCP and Peng brings significant improvement of quality assessment at the cost of a slightly increase of running time. The last row of RT (s) in Table IV presents the average running time of the images with 400×600 pixels processed by our method. Based on experiments, the running time of the image with $x \cdot y$ pixels is approximately $0.71 \cdot (x \cdot y)/(400 \cdot 600)$ seconds.

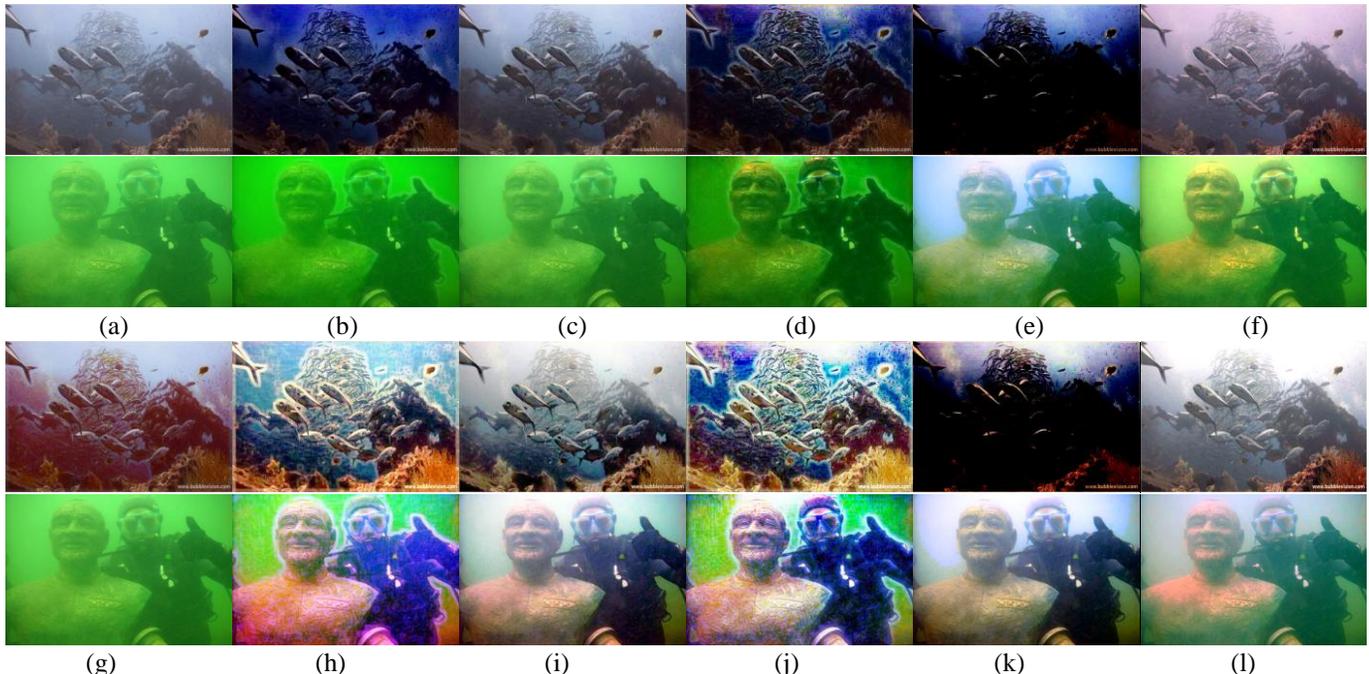

Fig. 16: Comparisons results. (a) Original images, (b) DCP, (c) MIP, (d) UDCP, (e) Peng, (f) OWCC, (g) Li, (h) DCP+HE, (i) MIP+HE, (j) UDCP+HE, (k) Peng+HE, (l) Ours.

The impact of color correction can be observed in Fig. 16 by comparing the top and bottom images with the corresponding methods. These IFM-based methods that restore the underwater images often fail to remove the green-bluish color. In particular, a greenish color still exists in the restored archaeology images. The color corrections have improved the brightness, color and contrast and effectively reduced the greenish. As shown in Fig. 16 (f), our restoration method can effectively reduce the effect of light absorption and scattering. However, the restored underwater images enhanced by HE are over-saturated as the image becomes too bright and unnatural, which also reduce the image valuable details, shown in Fig. 16 (h-j). The color correction may affect in a wrong way. As shown in Fig. 16 (k), the fish school is dimmer than the original image and even cannot be classified from the background scene.

Our results are shown in Fig. 16 (l), the final images enhanced by the improved color correction, considering the special characteristics of each R-G-B channel and the relation of RGB three color channels, are neither over-saturated nor over-enhanced, wherein the objects are better differentiated from the background. In sum, the effect of greenish illumination remains in the output images produced by physical-based methods. It is necessary to introduce a proper color correction to remove some blue-green illumination, and improve the color and saturation of the input image.

## V. DISCUSSION

We have proposed an underwater image enhancement method including underwater image restoration based on novel statistical models of BLs estimation and optimal TM estimation models and a simple color correction based on white balancing using the optimal gain factor. The accuracy of the estimated BLs and TMs influences on the restored quality of the underwater image based on the image formation model, meanwhile the accuracy of the BLs affects the estimation of the TMs. Hence an effective and efficient BL estimation model is the premise of successful TM estimation and underwater image restoration. In order to demonstrate that our proposed method can recover and enhance the quality of some different underwater images, and achieve similar results to those obtained using learning-based methods, we will further discuss the BLs, TMs and enhanced images with challenging underwater scenes and compare our proposed method with the restoration methods based on the CNN.

### A. Enhanced Images with Challenging Underwater Scenes

In order to show that our method can restore different underwater images under some challenging scenes, we select images under greenish scene, bluish scene, thickly hazed scene, low-visibility background scene, turbid scene and low-visibility scene as tested images, shown in the first row of Fig. 17. Some original challenging images, the corresponding estimated BL and enhanced images are shown in the Fig. 17. In the Fig. 17 (a-e), our statistical model of BL estimation can successfully estimate the global background light of the entire scene, avoiding the interference of the white block or bright point in the foreground object. Only we estimate the correct BL, the TMs can be deduced by the optical formation model. In the enhanced images of Fig. 17, the white balancing with optimal gain factor is essential to be introduced to improve the color and contrast of the dehazed images because the restoration method can only remove the haze and blur of the original image.

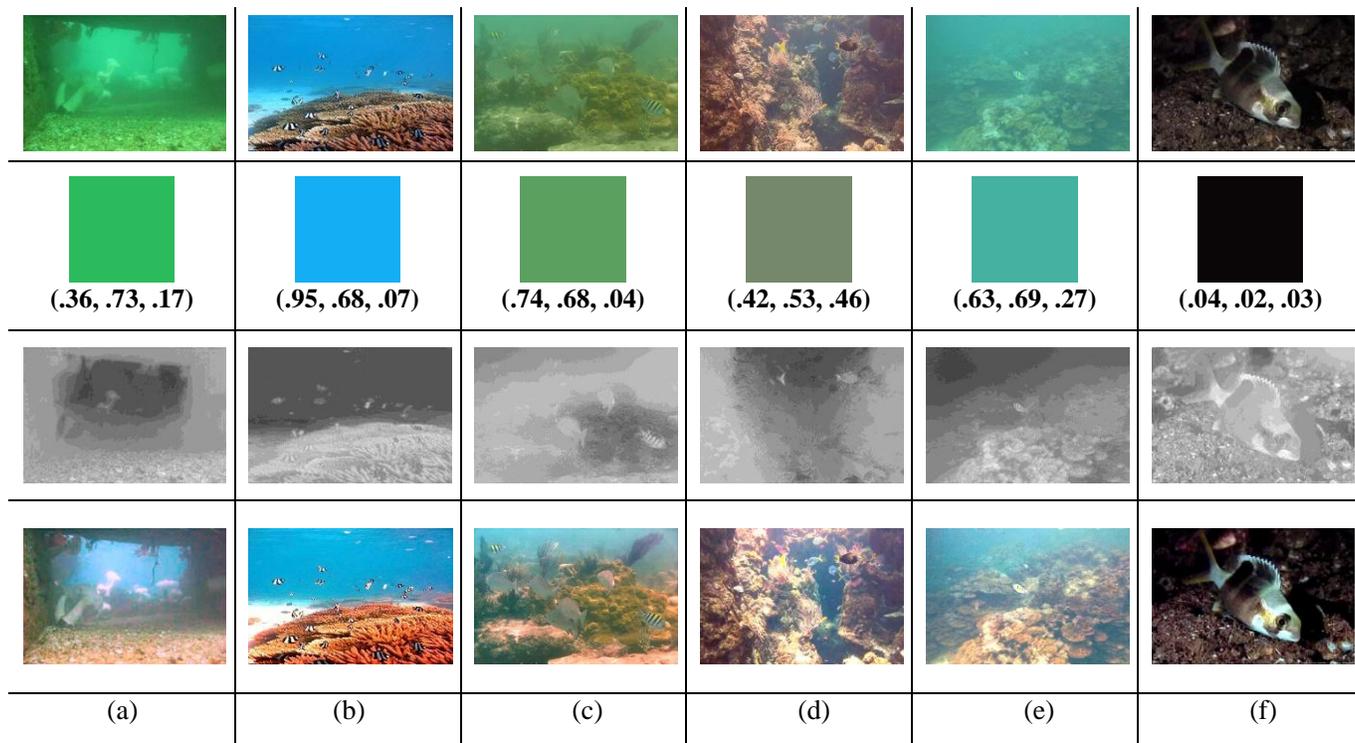

| | | | | | |
|---|---|---|---|---|---|
| (.36, .73, .17) | (.95, .68, .07) | (.74, .68, .04) | (.42, .53, .46) | (.63, .69, .27) | (.04, .02, .03) |

(a)    (b)    (c)    (d)    (e)    (f)

Fig. 17: Enhanced results. (a) Coarse image with greenish scene, (b) Coarse image with bluish scene, (c) Shoal of fish with thickly hazed scene, (d) Underwater chasm image with low-visibility background scene, (e) Turbid image with the severe distortion, (f) Fish image under a dim scene.

## B. Comparison with Restoration Methods based on the CNN

Reference [26] and our proposed method have different experimental conditions. The deep learning CNN needs to be trained with a huge amount of data, which is difficult for us to implement their work. To compare, we extracted some underwater images presented in [26] and give the results processed by our proposed method. Original images, Ding's results and our results are shown in Fig. 18. Although original images have the low resolution, form the compared results, we can spot that our proposed method with lower running time can be well-suitable for different underwater images enhancement in terms of higher contrast and visibility.

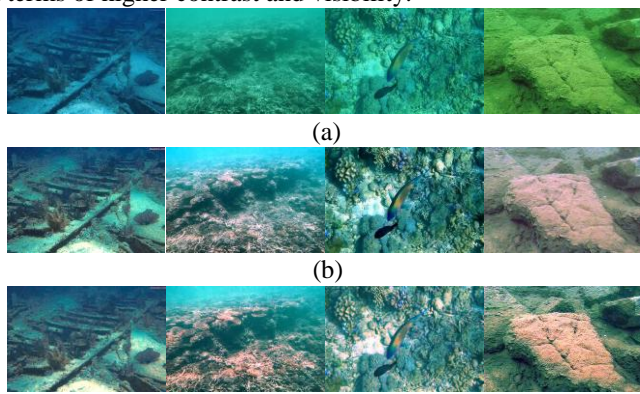

(a)

(b)

(c)

Fig. 18: Comparison results. (a) Original images, (b) Ding's method [37], (c) Ours.

## VI. CONCLUSION

We have proposed an underwater image enhancement method including underwater image restoration based on novel statistical models of BLs estimation and optimal TM estimation models, and a simple color correction based on improved white balance, in accordance with the characteristics of the underwater images. To guarantee the robustness and high accuracy of BL estimation model, we established a first MBALs database and made statistical analysis about the histogram distribution of R and G-B channel separately, referring to the MABLs. The TM of R channel was estimated based on the NUDCP, further compensated by the depth map and modified by the ARSM, and TMs of G-B channels were deducted according to the optical properties of the underwater image formation. Simple yet effective color correction was introduced to improve the contrast and color of the restored image.

Comprehensive evaluation approaches demonstrate the effectiveness of the MABLs, the high accuracy of our BLs estimation model, the rationality and low-complexity of our TMs estimation model and the superior performance of our proposed image enhancement model with other state-of-the-art image restoration or enhancement methods. Our proposed method has better results in processing various underwater images. Nevertheless, it was not possible to obtain high-quality BLs for some with close-shot objects, and it is partially also due also to the limitation of manual annotation of BLs based on the far scene. But it is very fortunate for these kinds of images not sensitive to the BLs in the process of restoration. The restored images using our method will not become further distorted than the original image. In future work, we will continue to optimize the statistical BL estimation model and TMs estimation model. We are working towards making our MABLs database available to the research community and the preliminary version of the database is available in the project website. Meanwhile, we are still extending the numbers of underwater images in our database and the corresponding MABLs, to train a more robust model of the background light estimation and make it public.